\documentclass[12pt,a4paper,twoside]{article}
\usepackage{amsmath}
\usepackage{amssymb}
\usepackage{amsfonts}
\usepackage{theorem}    
\usepackage{epsfig}   
\usepackage{helvet}

\newcommand{\R}{{\mathord{\mathbb R}}}
\newcommand{\Z}{{\mathord{\mathbb Z}}}
\newcommand{\N}{{\mathord{\mathbb N}}}
\newcommand{\C}{{\mathord{\mathbb C}}}

\newcommand{\T}{{\mathord{\mathbb T}}}

\newcommand{\mH}{{\mathcal H}}

\newcommand{\mL}{{\mathcal L}}
\newcommand{\mS}{{\mathcal S}}

\newcommand{\mE}{{\mathcal E}}

\newcommand{\mQ}{{\mathcal Q}}

\newcommand{\dd}{{\rm d}}
\newcommand{\e}{{\rm e}}
\newcommand{\ii}{{\rm i}}

\newcommand{\ran}{{\rm ran\,}}
\newcommand{\sign}{{\rm sign}}

\newcommand{\hh}{{{\mathfrak h}}}

\newcommand{\fF}{{\mathfrak F}}
\newcommand{\fA}{{\mathfrak A}}

\newcommand{\pf}{{\rm pf}}

\newcommand{\ind}{{\rm ind}}
\newcommand{\eps}{{\varepsilon}}
\newcommand{\dom}{{\rm dom\,}}

\newcommand{\rP}{{\rm P}}

\newcommand{\vi}{\varphi}
\newtheorem{thm}{Theorem}
\newtheorem{proposition}[thm]{Proposition}
\newtheorem{lemma}[thm]{Lemma}
\newtheorem{definition}[thm]{Definition}
{\theorembodyfont{\upshape} \newtheorem{remark}[thm]{\it Remark}}

\newtheorem{assumption}[thm]{Assumption}
\newtheorem{corollary}[thm]{Corollary}
\newcommand{\bd}{\begin{definition}}
\newcommand{\ed}{\end{definition}\vspace{1mm}}
\newcommand{\bes}{\begin{eqnarray*}}
\newcommand{\ees}{\end{eqnarray*}}
\newcommand{\be}{\begin{eqnarray}}
\newcommand{\ee}{\end{eqnarray}}
\newcommand{\bt}{\begin{thm}}\vspace{1mm}
\newcommand{\et}{\end{thm}}
\newcommand{\bc}{\begin{corollary}}
\newcommand{\ec}{\end{corollary}\vspace{1mm}}
\newcommand{\bl}{\begin{lemma}}
\newcommand{\el}{\end{lemma}\vspace{1mm}}
\newcommand{\bp}{\begin{proposition}}
\newcommand{\ep}{\end{proposition}\vspace{1mm}}
\newcommand{\br}{\begin{remark}}
\newcommand{\er}{\end{remark}\vspace{1mm}}
\newcommand{\ba}{\begin{assumption}}
\newcommand{\ea}{\end{assumption}\vspace{1mm}}
\newcommand{\bprf}{{\it Proof.}\, }
\newcommand{\eprf}{\hfill $\Box$ \vspace{5mm}}
\begin{document}
\pagestyle{myheadings}
\markboth{W. H. Aschbacher}{On the subleading order in the asymptotics of the nonequilibrium EFP}

\title{A remark on the subleading order in the asymptotics of the 
nonequilibrium emptiness formation probability}

\author{Walter H. Aschbacher
\footnote{walter.aschbacher@polytechnique.edu}
\footnote{Supported by the German Research Foundation (DFG).}
\\ \\
Ecole Polytechnique\\
Centre de Math\'ematiques Appliqu\'ees\\
UMR CNRS - 7641\\
91128 Palaiseau Cedex\\
France}

\date{}
\maketitle
\begin{abstract} 
We study the asymptotic behavior of the emptiness formation probability 
for large spin strings in a translation invariant  quasifree 
nonequilibrium steady state of the isotropic XY chain. 
Besides the overall exponential decay, we prove that, out of equilibrium, 
the exponent of the subleading power law contribution to the asymptotics 
is nonvanishing and strictly positive due to the singularities in the 
density of the steady state. 
\end{abstract}

\noindent {\it Keywords}\, Nonequilibrium steady state, emptiness 
formation probability, Toeplitz theory

\noindent{\it Mathematics Subject Classifications (2000)}\, 46L60, 
47B35, 82C10, 82C23
\section{Introduction}
\label{sec:introduction}
In this note, I propose to enlarge upon the study started in Aschbacher 
\cite{A07-1} of the asymptotic behavior of a special and important correlator, the so-called emptiness formation probability (EFP). Written down in the framework of a spin system over the two-sided discrete line, the EFP observable is given by
\be
A_n
=\prod_{i=1}^n p_i,
\ee
where $p_i$ denotes the projection $p:=(1-\sigma_3)/2$ at site $i$ onto the spin down configuration of the spin $\sigma_3$ in the $3$-direction. For a given state $\omega$ of the spin system, the probability that a ferromagnetic string of length $n$ is formed in this state is thus expressed by
\be
{\rm P}(n)
=\omega(A_n).
\ee

Due to the existence of the Jordan-Wigner transformation which maps,
in a certain sense, spins  onto fermions, the EFP has been heavily 
studied for states of the XY chain whose formal Hamiltonian is given
in Remark \ref{rem:XY} below. As a matter of fact,  this model becomes a gas of independent fermions under the Jordan-Wigner transformation, and it is thus  ideally suited for rigorous analysis.\footnote{Low-dimensional magnetic systems are also heavily studied  experimentally,
see for example Sologubenko et al. \cite{SGOVR01}.} 

The large $n$ behavior of the EFP in the XY chain has already been analyzed for the cases where
the state $\omega$ is a ground state or a thermal equilibrium state at positive 
temperature. In both cases, the EFP can be written as the determinant of the  section of a Toeplitz operator with scalar symbol.  Since the higher order asymptotics of a Toeplitz determinant is highly sensitive to the
regularity of the symbol of the Toeplitz operator, the asymptotic behavior of the ground state EFP is qualitatively different in the 
so-called  critical and noncritical regimes corresponding to
certain  values of the anisotropy and the exterior magnetic field of the XY chain.\footnote{I.e., in \eqref{H-XY} below, the parameters 
$\gamma$ and $\lambda$, respectively.} It has been found in Shiroishi et   
al. \cite{STN01} that the EFP decays
like a Gaussian in one of the critical regimes.\footnote{With some  additional explicit numerical prefactor and some power law prefactor, 
see Shiroishi et al. \cite{STN01} and references therein.} 
In a second critical regime and in all noncritical regimes, the EFP decays exponentially.\footnote{In contrast to the noncritical regime, there is an additional power law prefactor in the second critical regime whose exponent differs from the one in the first critical regime, see Abanov and Franchini \cite{AF03, FA05}.} 
These results have been derived by using well-known theorems of 
Szeg\H o, Widom, and Fisher-Hartwig, and the yet unproven Basor-Tracy conjecture and some of its extensions, see Widom \cite{W71} and
 B\"ottcher and Silbermann  \cite{BS99, BS06}. On the other hand, in thermal equilibrium at positive temperature,  the EFP can
again be shown to decay exponentially by using a theorem of Szeg\H o, 
see for example Shiroishi et al.  \cite{STN01} and  Franchini and Abanov \cite{FA05}.

In contrast, for the case where $\omega$ is the nonequilibrium steady state (NESS) constructed  in Aschbacher and Pillet \cite{AP03},\footnote{And in Araki and Ho 
\cite{ArakiHo00} for $\gamma=\lambda=0$.}
the EFP can still be written as a Toeplitz determinant, but now, the symbol is, in general, no longer scalar. Due to the lack of control of
higher order determinant asymptotics in  Toeplitz theory with
nontrivial irregular block symbols, I started off by studying  bounds on the leading asymptotic order for a class of general block Toeplitz
determinants in Aschbacher \cite{A07-1}.  It turned out that suitable basic spectral information on the density of the state is sufficient to derive a bound on the rate of the exponential decay of the EFP in general translation invariant fermionic quasifree states. 
This bound proved to be exact not only for the decay rates of the EFP in the ground states and the equilibrium states at positive temperature treated in  Abanov and Franchini \cite{AF03, FA05} and Shiroishi et al. \cite{STN01}, but it will also do so for the nonequilibrium situation
treated here exhibiting the so-called left mover-right mover structure already found in Aschbacher \cite{A07-2} and Aschbacher and Barbaroux  \cite{AB07} for nonequilibrium expectations of different correlation observables.\footnote{This has  already been noted in Aschbacher
  \cite{A07-1}.}
Hence, given this exponential decay in leading order which parallels qualitatively the behavior in thermal
equilibrium at positive temperature,
one may wonder whether there is some characteristic signature of the nonequilibrium left at some lower 
order of the  EFP asymptotics.  It turns out, and this is the main result of this note, that, in contradistinction to the leading order contribution, the subleading power law contribution to the large
$n$ asymptotics of the EFP in Fisher-Hartwig theory is sensitive to the singularity of the symbol of the underlying Toeplitz operator, and it has a strictly positive exponent if and only if the system is truly out of equilibrium. This may be interpreted as the manifestation, in subleading order, of the 
long-range nature of the underlying formal effective Hamiltonian of 
the NESS.\footnote{See Remark 3 in Aschbacher and Pillet \cite{AP03}. This effective Hamiltonian is to be understood on a formal level only. It has been shown by Matsui and Ogata \cite{MO} 
that there exists no dynamics on the Pauli spin algebra w.r.t. which
this NESS is a KMS state.}
This connection is not made precise here, though, but it is left to be studied in greater detail elsewhere.

Section \ref{sec:setting} contains the setting and Section 
\ref{sec:subleading} the main assertion.  The reader not familiar 
with quasifree states on CAR algebras and/or with Toeplitz theory 
may consult the Appendix where definitions and basic facts are 
collected.
\section{Nonequilibrium setting}
\label{sec:setting}

In this section, I shortly summarize the setting for the system
out of equilibrium used in Aschbacher and Pillet \cite{AP03}. In contradistinction to the presentation there, I will skip the formulation of the two-sided XY chain
as a spin system and rather focus directly on the underlying
$C^\ast$-dynamical system structure in terms of Bogoliubov automorphisms on a selfdual CAR algebra as introduced by Araki \cite{Araki84}.\footnote{For an introduction to the algebraic approach to open quantum systems, see also for example 
Aschbacher et al. \cite{AJPP06}.}

For some $N\in\{0\}\cup\N$, the nonequilibrium configuration is set up by cutting the  finite piece 
\be
\Z_\mS
:=\{x\in\Z\,|\, -N\le x\le N\}
\ee
out of the 
two-sided discrete line $\Z$.  This piece will play the role of the 
confined sample, whereas the remaining parts,
\be
\Z_L
&:=&\{x\in\Z\,|\, x\le -(N+1)\},\\
\Z_R
&:=&\{x\in\Z\,|\, x\ge N+1\},
\ee
will act as infinitely extended thermal reservoirs at different temperatures to which the sample will be suitably coupled.

We first specify the observables contained in the 
system to be considered.
\bd[Observables]
\label{def:obs}
Let $\fF(\hh)$ denote the fermionic Fock space built over the one-particle Hilbert space of wave functions on the discrete line,
\be
\hh
:=\ell^2(\Z).
\ee
With the help of the usual creation
and annihilation operators $a^\ast(f), a(f)\in \mL(\fF(\hh))$ for 
any $f\in\hh$,\footnote{The bounded operators on the Hilbert space $\mH$ are denoted
by $\mL(\mH)$.} the complex linear mapping 
$B: \hh^{\oplus 2}\to \mL(\fF(\hh))$ is defined, for 
$F:=[f_1,f_2]\in  \hh^{\oplus 2}$, by
\be
\label{B}
B(F)
:=a^\ast(f_1)+a(\bar f_2).
\ee
Moreover, the antiunitary involution 
$J:\hh^{\oplus 2}\to \hh^{\oplus 2}$ is given, for all $f_1, f_2\in\hh$,
by
\be
J[f_1,f_2]
:=[\bar f_2,\bar f_1].
\ee
The observables are described by the selfdual CAR algebra over 
$\hh^{\oplus 2}$ with involution 
$J$ generated by the operators $B(F)\in\mL(\fF(\hh))$ for all 
$F\in  \hh^{\oplus 2}$. I denote this algebra by  
$\fA:=\fA(\hh^{\oplus 2},J)$.\footnote{The concept of a selfdual CAR algebra has been introduced and developed by Araki 
\cite{Araki68, Araki71}. Here, it is just a convenient way of working with the linear combination \eqref{B}.}
\ed

The time evolution is generated as follows.
\bd[Dynamics]
\label{def:dynamics}
Let $\lambda\in\R$, and let $u\in\mL(\hh)$ be the translation operator defined by $(uf)(x):=f(x-1)$ for all $f\in\hh$ and all $x\in\Z$. The coupled and the decoupled one-particle Hamiltonians $h, h_0\in\mL(\hh)$ are defined by
\be
\label{h}
h
&:=&{\rm Re}(u)+\lambda,\\
h_0
&:=& h-(v_L+v_R),
\ee
respectively, where the decoupling operators  $v_L,v_R\in \mL^0(\hh)$ have the form
\be
v_L
&:=&{\rm Re}\big(u^{-(N+1)}p_0u^N\big),\\
v_R
&:=&{\rm Re}\big(u^Np_0u^{-(N+1)}\big),
\ee
and the projection $p_0\in\mL^0(\hh)$ is given by 
$p_0 f:=(\delta_0,f) \delta_0$ for all $f\in\hh$.\footnote{I write 
$\mL^0(\mH)$ for the finite rank operators on the Hilbert space $\mH$.
Moreover, $\delta_x\in\hh$ for $x\in\Z$ denotes the Kronecker function. Finally, for an operator $A\in\mL(\mH)$, the real part
is ${\rm Re}(A):=(A+A^\ast)/2$.} For all $t\in\R$, the coupled and the decoupled time evolutions are the Bogoliubov $\ast$-automorphisms $\tau^t, \tau_0^t
\in {\rm Aut}(\fA)$ defined on the generators $B(F)\in\fA$ with $F\in \hh^{\oplus 2}$ by
\be
\tau^t(B(F))
&:=& B(\e^{\ii t (h\oplus -h) }F),\\
\tau^t_0(B(F))
&:=& B(\e^{\ii t (h_0\oplus -h_0)}F).
\ee
\ed
\br
\label{rem:XY}
As mentioned above, this model has its origin in the XY spin chain 
whose formal Hamiltonian is given by 
\be
\label{H-XY}
H=-\frac{1}{4}\sum_{x\in\Z}\left\{(1+\gamma)\,\sigma_1^{(x)}\sigma_1^{(x+1)}+(1-\gamma)\,\sigma_2^{(x)}\sigma_2^{(x+1)}+2\lambda
 \, \sigma_3^{(x)}\right\},
\ee
where $\gamma\in(-1,1)$ denotes the anisotropy, $\lambda\in\R$
the external magnetic field, and the Pauli basis
of $\C^{2\times 2}$ reads
\be
\label{Pauli} 
\sigma_0
=\left[\begin{array}{cc}
1 & 0
\\ 0&1
\end{array}\right],\quad 
\sigma_1
=\left[\begin{array}{cc}
0 & 1\\
1& 0
\end{array}\right],\quad 
\sigma_2
=\left[\begin{array}{cc}
0 &-\ii\\ 
\ii & 0
\end{array}\right],\quad
\sigma_3
=\left[\begin{array}{cc}
1 & 0
\\ 0& -1
\end{array}\right].
\ee
The Hamiltonian $h$ from \eqref{h} corresponds to  the 
isotropic XY chain, i.e. to the case where $\gamma=0$.
\er

The left and right reservoirs carry the inverse temperatures $\beta_L$ and $\beta_R$, respectively. Pour fixer les id\'ees, we assume
w.l.o.g. that they satisfy
\be
0
<\beta_L
\le\beta_R
<\infty.
\ee
We next specify the state in which the system is prepared initially.
It consists of a KMS state at the corresponding temperature for each
reservoir, and, w.l.o.g., of the chaotic state for the sample. For the definition of quasifree states, see Appendix \ref{app:qf}.
\bd[Initial state]
\label{def:initial}
The initial state 
$\omega_0\in\mQ(\fA)$ is the quasifree state specified by the density
$S_0\in\mL(\hh^{\oplus 2})$ of the form
\be
S_0
:= s_{0,-} \oplus s_{0,+},
\ee
where the operators $s_{0,\pm}\in\mL(\hh)$ are defined by
\be
s_{0,\pm}
:=(1+\e^{\pm k_0})^{-1},
\ee
and $k_0\in\mL(\hh\simeq\hh_L\oplus\hh_\mS\oplus\hh_R)$ is given by
\be
k_0
:=\beta_L h_{L}\oplus 0 \oplus \beta_R h_{R}.
\ee
Here, for $\alpha=L,\mS, R$, I used the definitions $\hh_\alpha:=\ell^2(\Z_\alpha)$ and $h_{\alpha}:=i_\alpha^\ast h i_\alpha$,  where 
$i_\alpha: \hh_\alpha\to\hh$ is the natural injection defined, for any 
$f\in\hh_\alpha$,  by
$i_\alpha(\{f(y)\}_{y\in\Z_\alpha})(x)
:=f(x)$ if $x\in\Z_\alpha$, and zero otherwise.
\ed

The following definition is due to  Ruelle \cite{Ruelle00}.
\bd[NESS]
A NESS associated with the $C^\ast$-dynamical system $(\fA,\tau)$ 
having the initial state $\omega_0\in\mE(\fA)$ is a weak-$\ast$ limit 
point for $T\to\infty$ of the net
\be
\left\{\frac1T\int_0^T\dd t\,\, \omega_0\circ \tau^t\,\,\Big|\,\, 
T>0\right\}.
\ee
\ed

In the model specified by the Definitions \ref{def:obs}, \ref{def:dynamics},
and \ref{def:initial},  we get the following NESS using
the scattering approach of Ruelle \cite{Ruelle00}.
\bt[XY NESS]
\label{thm:ness}
There exists a unique quasifree NESS $\omega\in \mQ(\fA)$ w.r.t. the initial state $\omega_0\in \mQ(\fA)$
and the coupled dynamics $\tau^t\in{\rm Aut}(\fA)$
whose density $S\in\mL(\mH)$ reads
\be
S
=s_-\oplus s_+,
\ee
where the operators $\hat s_\pm \in\mL(\hat \hh)$ act in momentum space 
$\hat\hh:= L^2(\T)$ as multiplication by 
\be
\label{hat-s}
\hat s_\pm(\e^{\ii k})
:=\frac12\,(1\pm \varrho_\pm(\e^{\ii k})),
\ee
and the functions $\varrho_\pm:\T\to(-1,1)$ are defined by
\be
\label{rho}
\varrho_\pm(\e^{\ii k}):=
\tanh\!\big[\tfrac12 (\beta\pm\sign(\sin k)\delta) (\lambda+\cos k)\big].
\ee
Here, we set $\beta:=(\beta_R+\beta_L)/2$ and 
$\delta:=(\beta_R-\beta_L)/2$, and the sign function 
$\sign:\R\to\{\pm1\}$ is defined by $\sign(x):=1$ if $x\ge 0$, and $\sign(x):=-1$ if $x<0$.
\et
\bprf
See Aschbacher and Pillet \cite{AP03}.
\eprf

The main object of our study is the following.
\bd[NESS EFP]
\label{def:FF}
Let $n\in\N$. The EFP observable $A_n\in\fA$ is defined 
by
\be
A_n
:= \prod_{i=1}^{2n} B(F_i),
\ee
where, for all $i\in\N$, the form factors $F_i\in\hh^{\oplus 2}$ are given 
by
\be
F_{2i-1}
&:=& u^i\oplus u^i\, G_1,\\
F_{2i}
&:=& u^i\oplus u^i\, G_2,
\ee
and the initial form factors $G_1, G_2\in \hh^{\oplus 2}$ 
look like 
\be
G_1
:=JG_2
:=[0,\delta_0].
\ee
Moreover, the expectation value 
$\rP:\N\to[0,1]$ of the EFP observable $A_n\in\fA$ in the NESS 
$\omega\in\mE(\fA)$ is denoted by\footnote{As for the name EFP, note that  
$A_n
=\prod_{i=1}^n a_i a^\ast_i$, and that, for $B_n:=\prod_{i=1}^n a_i$,
we have
\bes
0
\le\rP(n)
=\omega(B_nB_n^\ast)
\le {\|B_n\|}^2
\le \prod_{i=1}^n{\|\delta_i\|}^2
= 1.
\ees
}
\be
\rP(n)
:=\omega(A_n).
\ee
\ed

The next assertion states the main structural property of the EFP
correlation function. For the basic facts of Toeplitz theory, see Appendix \ref{app:Toeplitz}.
\bp[EFP determinantal structure]
\label{prop:symbol}
The NESS EFP $\rP:\N\to [0,1]$  is given by the  determinant of the finite 
section of the Toeplitz operator $T[\hat s_-]\in\mL(\ell^2(\N))$,
\be
\label{PT}
\rP(n)
=\det (T_n[\hat s_-]).
\ee
\ep
\bprf
Proceeding as in Aschbacher and Barbaroux \cite{AB07}, we have that, on one hand,
the skew-symmetric EFP correlation matrix 
$\Omega_n\in\R^{2n\times 2n}$, defined, for
$i,j=1,...,2n$, by
\be
\Omega_{n,ij}
=\begin{cases}
\omega(B(F_i)B(F_j)),  & \mbox{if\, $i<j$},\\ 
0,                     & \mbox{if\, $i=j$},\\
-\,\omega(B(F_j)B(F_i)), & \mbox{if\, $i>j$},
\end{cases}
\ee
where $F_i\in\hh^{\oplus 2}$ for $i\in\N$ are the form factors from
Definition \ref{def:FF}, relates to the EFP as
\be
\rP(n)
=\pf (\Omega_n),
\ee
and, on the other hand, that it has the Toeplitz structure
\be
\Omega_n
=T_n[a_\rP].
\ee
Here, $a_\rP\in L^\infty_{2\times 2}(\T)$ is the block symbol of the Toeplitz operator $T[a_\rP]\in\mL(\ell_2^2(\N))$ which I computed in 
Aschbacher
\cite{A07-1} to be of the form $a_\rP=(\hat s-p)\sigma_1$, where 
$\hat s=\hat s_-\oplus\hat s_+$ is the density of the NESS 
$\omega\in\mQ(\fA)$  in momentum space and $p=(1-\sigma_3)/2$.  
Theorem
\ref{thm:ness} then implies that, in the present case, the symbol has the form
\be
a_\rP
=\begin{bmatrix}
0 & \hat s_- \\
\hat s_+ -1 & 0
\end{bmatrix}.
\ee
Hence, there exists an $R\in O(2n)$ with $\det(R)=(-1)^{n(n-1)/2}$ s.t., using Lemma \ref{lem:pfaffian}, we can reduce the 
block Toeplitz Pfaffian to a scalar Toeplitz determinant,
\be
\rP(n)
&=& \pf (T_n[a_\rP])\nonumber\\
&=& (-1)^{\frac{n(n-1)}{2}} \,
\pf\bigg(\!\!
\begin{bmatrix}
0 &  T_n[\hat s_-]\\
T_n[\hat s_+ -1] & 0
\end{bmatrix}
\!\!\bigg)\nonumber\\
&=& \det (T_n[\hat s_-]),
\ee
where I used the fact that 
$T_n[\hat s_+ -1]=-{T_n[\hat s_-]}^{\rm t}$.\footnote{$O(2n)$ stands for the orthogonal matrices in 
$\R^{2n\times 2n}$.}
This is the assertion.
\eprf

\section{Subleading order in the NESS EFP asymptotics}
\label{sec:subleading}
\begin{figure}
\centering
\rotatebox{-90}{
\includegraphics[width=7cm,height=9cm]{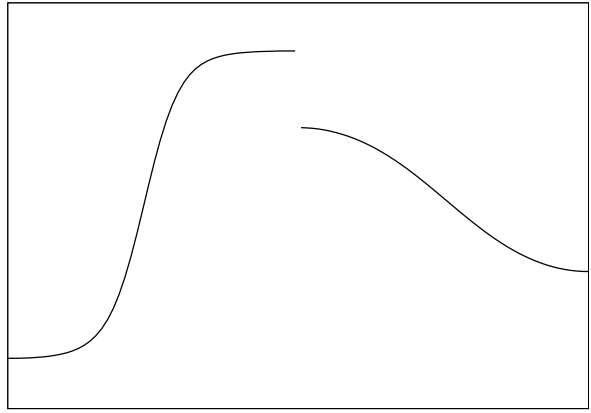} 
}
\caption{The symbol $\hat s_-(\e^{\ii k})$ with $k\in(-\pi,\pi]$ for 
$\beta=\tfrac32$, $\delta=1$, and 
$\lambda=\tfrac{1}{10}$.}
\label{fig:gnu-symbol}
\end{figure}

Due to Proposition \ref{prop:symbol}, the study of the large $n$ behavior of the EFP correlation function boils down to the analysis
of a large truncated Toeplitz operator whose symbol is scalar and has 
the form given in Theorem \ref{thm:ness},
\be
\label{s-}
\hat s_-(\e^{\ii k})
=\frac12 \left(1- \tanh\!\left[\tfrac12 (\beta-\sign(\sin k)\delta) (\lambda+\cos k)\right]\right),
\ee
see Figure \ref{fig:gnu-symbol}.
In a true nonequilibrium situation, i.e. for $\delta>0$, the r.h.s.
of \eqref{s-} is no longer continuous. Hence, as described in the 
introduction,  we want to study the asymptotic behavior of the EFP NESS 
with the help of the so-called Fisher-Hartwig
theory whose main content is summarized in Theorem \ref{thm:FH} of 
Appendix \ref{app:FH}. We first introduce the so-called pure jump
symbols. For notation and definitions, see Appendix
\ref{app:fc}.
\bd[Pure jump] 
\label{def:jump}
Let the argument function $\arg: \C\setminus\{0\}\to \R$ be defined by 
$z=:|z|\,\e^{\ii \arg(z)}$ and $\arg(z)\in (-\pi,\pi]$ for all 
$z\in \C\setminus\{0\}$. For $\beta_0\in\C$ and $t_0\in\T$, the pure jump 
symbol $\varphi_{\beta_0,t_0}\in PC_0(\T)$ is defined, for all $t\in\T$, by
\begin{eqnarray}
\label{jump}
\vi_{\beta_0,t_0}(t)
:=\e^{\ii\beta_0\arg\left(-\frac{t}{t_0}\right)}.
\end{eqnarray}
\ed
\br
Note that $\varphi_{\beta_0,t_0}\in PC_0(\T)$ has at most one jump 
discontinuity at the point $t_0$, namely 
\be
\varphi_{\beta_0,t_0}(t_0\pm 0)
=\e^{\mp\ii\pi\beta_0}.
\ee
\er

Moreover, the so-called jump phases are defined as follows.
\bd[Jump phases]
\label{lem:phase}
Let $a\in PC_0(\T)$ with $\Lambda_a=\{t_j\in\T\,|\, j=1,...,m\}$, and let 
$a(t_j\pm 0)\neq 0$ for $j=1,...,m$. The numbers 
$\beta_j\in\C$ for $j=1,...,m$, called the pure jump phases, are defined by
\be
\label{phase-1}
\frac{a(t_j-0)}{a(t_j+0)}
=\e^{2\pi \ii\beta_j}.
\ee
\ed

\begin{figure}
\centering
\rotatebox{-90}{
\includegraphics[width=7cm,height=9cm]{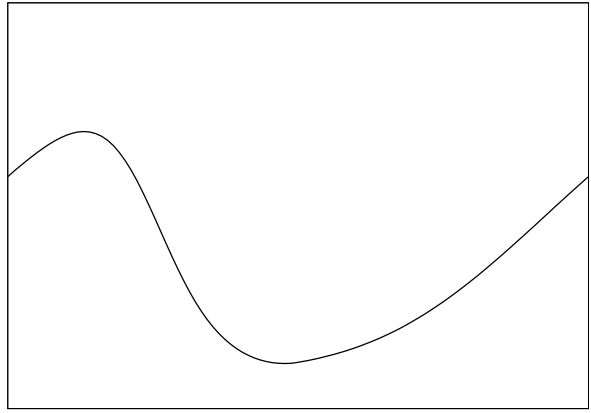} 
}
\caption{The regularized symbol $b_\rP(\e^{\ii k})$ with $k\in(-\pi,\pi]$ 
for $\beta=\tfrac32$, $\delta=1$, and $\lambda=\tfrac{1}{10}$.}
\label{fig:gnu-regsymbol}
\end{figure}

Next, we define a regularized symbol which will be extracted from 
$\hat s_-$ below.
\bd[Regularized symbol]
\label{def:regsym}
Let $t_1:=1$ and $t_2:=-1$. The regularized symbol $b_\rP\in C(\T)$ is 
defined by
\be
\label{b}
b_\rP(\e^{\ii k})
:= 
\left(\frac{\tau_L(t_1)}{\tau_R(t_1)}
\frac{\tau_R(t_2)}{\tau_L(t_2)}\right)^{\frac{k}{2\pi}} 
\begin{cases}
\sqrt{\frac{\tau_R(t_1)}{\tau_L(t_1)}}\,  
\tau_L(\e^{\ii k}), & \mbox{if \, $0\le k \le \pi$},\\
\sqrt{\frac{\tau_L(t_1)}{\tau_R(t_1)}}\,  
\tau_R(\e^{\ii k}), & \mbox{if \,$-\pi < k < 0$},
\end{cases}
\ee
where, for $\alpha=L,R$, the function $\tau_\alpha:\T\to (0,1)$ has the 
form
\be
\tau_\alpha(\e^{\ii k})
:=\frac12
\left(1-\tanh\big[\tfrac12 \beta_\alpha(\lambda+\cos k)\big]\right),
\ee
see Figure \ref{fig:gnu-regsymbol}.
\ed

Using the pure jump phases and the regularized symbol, we can recast
$\hat s_-$ into the following form.
\bl[Restricted Fisher-Hartwig form]
\label{lem:symbol}
The NESS EFP symbol $\hat s_- \in L^\infty(\T)$ has the following 
properties.
\begin{enumerate}
\item [(a)] $\hat s_-\in PC_0(\T)$
\item [(b)] $\Lambda_{\hat s_-}\!=\{t_1,t_2\}$ 
\item [(c)] The jump phases of $\hat s_-$ at the points 
$t_1$ and $t_2$ are given by
\be
\label{beta1}
\beta_1
&=&-\frac{\ii}{2\pi}\log\left(\frac{\tau_R(t_1)}{\tau_L(t_1)}\right),\\
\label{beta2}
\beta_2
&=&-\frac{\ii}{2\pi}\log\left(\frac{\tau_L(t_2)}{\tau_R(t_2)}\right).
\ee
\item [(d)] The symbol $\hat s_-\in L^\infty(\T)$ can be written as 
\be
\label{symbol}
\hat s_-
= b_\rP \vi_{\beta_1,t_1} \vi_{\beta_2,t_2}.
\ee
\end{enumerate}
\el
\bprf
The assertions (a) and (b) immediately follow
from the form of the symbol $\hat s_- \in L^\infty(\T)$ given in 
\eqref{s-}. Moreover, using the choice 
\be
\label{phase-2}
\beta_j
=\frac{1}{2\pi}\, \arg\bigg(\frac{\hat s_-(t_j-0)}{\hat s_-(t_j+0)}\bigg)
-\frac{\ii}{2\pi}\, \log\bigg(\frac{\hat s_-(t_j-0)}{\hat s_-(t_j+0)}
\bigg),
\ee
where the argument function is given in Definition 
\ref{def:jump}, we get the pure jump phases \eqref{beta1} and 
\eqref{beta2} in assertion (c). As for
assertion (d), writing, for $k\in (-\pi,\pi]$, 
\be
\vi_{\beta_1,t_1}(\e^{\ii k})
&=& \left(\frac{\tau_R(t_1)}{\tau_L(t_1)}\right)
^{\frac{k+\sign(-k)\pi}{2\pi}},\\
\vi_{\beta_2,t_2}(\e^{\ii k})
&=& \left(\frac{\tau_L(t_2)}{\tau_R(t_2)}\right)^{\frac{k}{2\pi}},
\ee
with the  sign function defined after \eqref{rho},
we get equality \eqref{symbol} in $L^\infty(\T)$ involving
the regularized symbol $b_\rP\in C(\T)$.
\eprf

In order to be able to apply the Fisher-Hartwig theory to the symbol
$\hat s_- $, we have to make sure that the regularized symbol 
$b_\rP\in C(\T)$ is indeed sufficiently regular. 
\bl[Fisher-Hartwig regularity]
\label{lem:b}
The regularized symbol $b_\rP\in C(\T)$ has the following properties.
\begin{enumerate}
\item[(a)] $b_\rP(t)\neq 0$ for all $t\in\T$
\item[(b)] $\ind(b_\rP)=0$
\item[(c)] $b_\rP\in B^1_1(\T)$
\end{enumerate}
\el
\bprf
It follows from \eqref{b} that $b_\rP(t)>0$ for all $t\in\T$
which implies assertions (a) and (b). As for assertion (c), we use 
the fact given in Lemma \ref{lem:regsym2} (b) of Appendix \ref{app:FH}  
that $b_\rP\in C^1(\T)$. This allows us to bound the integrand in 
\eqref{Besov} in the Definition \ref{def:Besov} of the Besov space 
$B_1^1(\T)$ as
\be
\label{est1}
\int_{-\pi}^\pi\frac{\dd k}{k^2}\,\, {\|\Delta^2_k b_\rP\|}_{L^1(\T)} 
&=& \int_{-\pi}^\pi\frac{\dd k}{k^2}\int_{-\pi}^\pi\dd \theta\,\,
|b_\rP(\e^{\ii(\theta+k)})-2b_
P(\e^{\ii \theta})+b_\rP(\e^{\ii(\theta-k)})|\nonumber\\
&\le&\int_{-\pi}^\pi\frac{\dd k}{|k|}\int_{-\pi}^\pi\dd \theta
\int_0^1\dd t\,\,|b'_\rP(\e^{\ii(\theta+tk)})-b'_\rP(\e^{\ii(\theta-tk)})|.
\ee
Since, due to Lemma \ref{lem:regsym2} (a)--(c), $b'_\rP$ is continuous 
and differentiable at all but finitely many points having a bounded 
derivative ${\|b_\rP''\|}_{L^\infty(\T)}<\infty$, we have
$b'_\rP\in AC(\T)$, and, thus, it follows from the fundamental theorem of 
calculus that $b'_\rP\in {\rm Lip}(\T)$ with Lipschitz constant 
${\|b_\rP''\|}_{L^\infty(\T)}$. Using the Lipschitz continuity on the 
r.h.s. of \eqref{est1}, we get
\be
\label{est2}
\int_{-\pi}^\pi\frac{\dd k}{k^2}\,\, {\|\Delta^2_k b_\rP\|}_{L^1(\T)} 
\le 4\pi^2 {\|b_\rP''\|}_{L^\infty(\T)}
<\infty.
\ee
Hence, we arrive at assertion (c).
\eprf

We are now ready to formulate the main result of this note.
\bt[NESS EFP asymptotics]
\label{thm:EFP-main}
For $n\to\infty$, the NESS EFP  behaves asymptotically as
\be
\label{EFP-main}
\rP(n)
\sim {G(b_\rP)}^n n^{Q_\rP} F(\hat s_-),
\ee
where the base of the exponential factor is given by
\be
\label{GbP}
G(b_\rP)
= \exp\!\left(\frac12 \!
\sum_{\alpha=L,R} \int_{-\pi}^\pi\frac{\dd k}{2\pi}\,\,
\log (\tau_\alpha(\e^{\ii k}))\right),
\ee
satisfying $0<G(b_\rP)<1$ for all inverse temperatures in the range 
$0<\beta_L\le\beta_R<\infty$. Furthermore, the exponent of the power law factor has the form
\be
\label{QP}
Q_\rP
= \frac{1}{4\pi^2}\sum_{j=1,2}
\left(\log\left(\frac{\tau_R(t_j)}{\tau_L(t_j)}\right)\right)^{\!\!2}.
\ee
Thus,  $Q_\rP>0$ if and only if $\beta_L\neq\beta_R$.
\et
\bprf
Due to \eqref{symbol} of Lemma \ref{lem:symbol}, the nonequilibrium symbol $\hat s_-$ has the form  
$\hat s_-=b_\rP\vi_{\beta_1,t_1}\vi_{\beta_2,t_2}$, where 
$t_1\neq t_2$ and, w.r.t. to the form \eqref{FH-symbol} of the restricted Fisher-Hartwig symbol, we have
$\alpha_1=\alpha_2=0$,  and ${\rm Re}(\beta_1)={\rm Re}(\beta_2)=0$ from
\eqref{beta1} and \eqref{beta2} in Lemma \ref{lem:symbol}. Hence, assumptions (a) and (b) of Theorem \ref{thm:FH} from Appendix 
\ref{app:FH} are satisfied. Moreover,
due to Lemma \ref{lem:b}, assumptions (c)--(d) of Theorem \ref{thm:FH}
are also satisfied by the regularized symbol. Then, \eqref{EFP-main}
follows from \eqref{FH} in Theorem \ref{thm:FH}, and it remains to
derive the  exponential, the power law, and the constant factors 
in \eqref{EFP-main} with the help of \eqref{Gb}--\eqref{Fa}. 
As for $G(b_\rP)$, we get \eqref{GbP} from \eqref{Gb} and \eqref{b} in Definition \ref{def:regsym}. Moreover, plugging \eqref{beta1} and 
\eqref{beta2} into \eqref{Q}, we find \eqref{QP}. Finally, using \eqref{Fa}, the last factor on the r.h.s. of \eqref{EFP-main} has the form
\be
F(\hat s_-)
=E(b_\rP)\, 2^{2\beta_1\beta_2} \prod_{j=1,2}
\left(\frac{b_{\rP,+}(t_j)}{b_{\rP,-}(t_j)}\right)^{\beta_j}
\prod_{j=1,2} {\rm G}(1+\beta_j){\rm G}(1-\beta_j),
\ee
where $E(b_\rP)$, $b_{\rP,\pm}(t_j)$, and ${\rm G}(1\pm\beta_j)$
are given in \eqref{Eb}, \eqref{bpm}, and \eqref{G}, respectively.
\eprf
\br
In Aschbacher \cite{A07-1}, I derived a bound on the decay rate of the 
exponential decay for the NESS EFP in the more general anisotropic XY chain. As noted there and discussed in the present introduction, Theorem \ref{thm:EFP-main} yields that this bound is exact for the special isotropic case at hand.\footnote{This can also be seen
by directly using Szeg\H o's first limit theorem, see for example
B\"ottcher and Silbermann \cite[p.139]{BS99}.}
\er

\appendix  
\section{Fermionic quasifree states}  
\label{app:qf}  

Let $\fA$ be the selfdual CAR algebra from Definition \ref{def:obs}.
We denote by $\mE(\fA)$ the set of states on the $C^\ast$ algebra 
$\fA$.\footnote{I.e. the normalized positive linear functionals on 
$\fA$.} 
\bd[Density]
The density of a state $\omega\in\mE(\fA)$ is defined to be the
operator $S\in\mL(\hh^{\oplus 2})$ with $0\le S^\ast=S\le 1$ and 
$JSJ=1-S$ satisfying, for all $F,G\in \hh^{\oplus 2}$,
\be
\omega(B^\ast(F)B(G))
=(F,SG).
\ee
\ed 

A special class of states are the important fermionic quasifree states.
\bd[Quasifree state]
A state $\omega\in\mE(\fA)$ is called quasifree 
if it vanishes on the odd polynomials in the generators, and if it
is a Pfaffian
on the even polynomials in the generators, i.e. if, for all 
$F_1,...,F_{2n}\in \hh^{\oplus 2}$ and for any $n\in\N$, we have
\be
\omega(B(F_1)...B(F_{2n}))
=\pf(\Omega_n),
\ee
where the skew-symmetric  matrix 
$\Omega_n\in\C_{ss}^{2n\times 2n}$ is  defined, for
$i,j=1,...,2n$, by
\be
\Omega_{n,ij}
:=\begin{cases}
\omega(B(F_i)B(F_j)),  & \mbox{if\, $i<j$},\\ 
0,                     & \mbox{if\, $i=j$},\\
-\omega(B(F_j)B(F_i)), & \mbox{if\, $i>j$}.
\end{cases}
\ee
Here, the Pfaffian $\pf: \C_{ss}^{2n\times 2n}\to\C$ is defined, on all
skew-symmetric matrices $A\in \C_{ss}^{2n\times 2n}:=\{A\in\C^{2n\times 2n}\,|\, A^{\rm t}=-A\}$,\footnote{$A^{\rm t}$ is the transposition of the matrix 
$A\in\C^{n\times n}$.} by
\be
\pf(A)
:=\sum_{\pi}\sign(\pi)\prod_{j=1}^nA_{\pi(2j-1),\pi(2j)},
\ee
where the sum is running over all pairings of the set $\{1,2,...,2n\}$,
i.e. over all the $(2n)!/(2^n n!)$ permutations $\pi$ in the permutation group of $2n$ elements which satisfy $\pi(2j-1)<\pi(2j+1)$ and $\pi(2j-1)<\pi(2j)$, see Figure \ref{fig:pairings}. 
The set of quasifree states is denoted by $\mQ(\fA)$.
\ed

\vspace{-1.8cm}

\begin{center}
\begin{figure}[h!]
\setlength{\unitlength}{0.9cm}
\begin{center}
\begin{picture}(22,2)
\multiput(1,0)(0.5,0){6}{\circle*{0.15}}
\put(1.25,0){\oval(0.5,1.5)[t]}
\put(2.25,0){\oval(0.5,1.5)[t]}
\put(3.25,0){\oval(0.5,1.5)[t]}
\put(1.05,-0.6){$\pi=(123456)$}
\put(3.8,0.25){$-$}
\multiput(4.5,0)(0.5,0){6}{\circle*{0.15}}
\put(4.75,0){\oval(0.5,1.5)[t]}
\put(6,0){\oval(1,1.5)[t]}
\put(6.5,0){\oval(1,1.5)[t]}
\put(4.55,-0.6){$\pi=(123546)$}
\put(7.3,0.25){$+$}
\multiput(8,0)(0.5,0){6}{\circle*{0.15}}
\put(8.25,0){\oval(0.5,1.5)[t]}
\put(9.75,0){\oval(1.5,1.5)[t]}
\put(9.75,0){\oval(0.5,1.5)[t]}
\put(8.05,-0.6){$\pi=(123645)$}
\put(10.8,0.25){$-$}
\multiput(11.5,0)(0.5,0){6}{\circle*{0.15}}
\put(12,0){\oval(1,1.5)[t]}
\put(12.5,0){\oval(1,1.5)[t]}
\put(13.75,0){\oval(0.5,1.5)[t]}
\put(11.55,-0.6){$\pi=(132456)$}
\put(14.3,0.25){$+$}
\put(14.8,0.25){\ldots}
\end{picture}
\end{center}
\vspace{0.5cm}
\caption{\label{fig:pairings} Some of the pairings for $n=3$. The total number of intersections  $I$ relates to the signature as $\sign(\pi)=(-1)^I$.}
\end{figure}
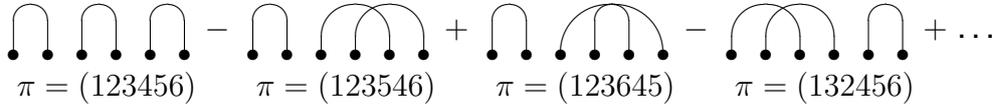
\end{center}

The following lemma has been used in Section \ref{sec:setting}.
\bl[Pfaffian]
The Pfaffian has the following properties.
\begin{enumerate}
\item[(a)] Let $X,Y\in\C^{2n\times 2n}$ with $Y^{\rm t}=-Y$. Then, 
\label{lem:pfaffian}
\be
\pf (XYX^{\rm t})
=\det(X)\,\pf(Y).
\ee
\item[(b)] Let $X\in\C^{n\times n}$. Then, 
\be
\pf \left(
\begin{bmatrix}
0 &  X\\
-X^{\rm t} & 0
\end{bmatrix}
\right)
=(-1)^{\frac{n(n-1)}{2}}\,\det(X).
\ee
\end{enumerate}
\el
\bprf
See for example Stembridge \cite{Stembridge90}.
\eprf

\section{Toeplitz theory}  
\label{app:Toeplitz}  
The material of this section is taken from B\"ottcher and Silbermann 
\cite{BS99, BS06}.
\subsection{Function classes} 
\label{app:fc}
Let $\T:=\{t\in\C\,|\,|t|=1\}$ stand for the unit circle. We denote by 
$C(\T)$ the continuous functions, by $C^m(\T)$ the $m$ times 
continuously differentiable functions, and by $L^p(\T)$ with 
$1\le p\le\infty$ the usual Lebesgue spaces. Moreover,
$AC(\T)$ and ${\rm Lip}(\T)$ stand for the absolutely continuous 
and the Lipschitz continuous functions on $\T$.
Finally, we need the following function class. 
\bd[Piecewise continuous]
The set of piecewise continuous functions is defined by
\be
PC(\T)
:=\{a\in L^\infty(\T)\,|\, \mbox{The limits $\lim_{\eps\to 0^+} 
a(\e^{\ii (k\pm \eps)})$ exist for all $k\in (-\pi,\pi]$}\}.
\ee
For  $a\in PC(\T)$ and any $t\in\T$ of the form 
$t=\e^{\ii k}$ with $k\in (-\pi,\pi]$, we use the notation
\be
a(t\pm 0)
:=\lim_{\eps\to 0^+} a(\e^{\ii (k\pm \eps)}).
\ee
Moreover, the set of jumps of $a\in PC(\T)$ is defined by
\be
\Lambda_a
:=\{t\in\T\,|\, a(t-0)\neq a(t+0)\}.
\ee
Finally, the set of piecewise continuous functions with finitely many 
jumps is defined by
\be
PC_0(\T)
:=\{a\in PC(\T)\,|\, {\rm card}(\Lambda_a)<\infty\}.
\ee
\ed
In order to be able to make use of the Fisher-Hartwig theory from 
Appendix \ref{app:FH}, we also need to introduce the following 
function class.
\bd[Besov space]
\label{def:Besov}
Let  $1\le p<\infty$ and $k\in (-\pi,\pi]$. The operator
$\Delta_k: L^p(\T)\to L^p(\T)$ is defined, on all $f\in L^p(\T)$, 
and for all $\theta\in (-\pi,\pi]$, by
\be
(\Delta_k f)(\e^{\ii \theta})
:=f(\e^{\ii(\theta+k)})-f(\e^{\ii \theta}).
\ee
Moreover, for any $n\in\N$, we recursively set 
$\Delta_k^n:=\Delta_k\Delta_k^{n-1}$. For $\alpha>0$ and $1\le p<\infty$, 
the Besov class is defined by
\begin{eqnarray}
\label{Besov}
B^\alpha_p(\T)
:=\Big\{f\in L^p(\T)\, \Big|\, \int_{-\pi}^\pi \dd k\,\,
|k|^{-(1+\alpha p)}\, {\|\Delta_k^nf\|}^p_{L^p(\T)}<\infty\Big\},
\end{eqnarray}
where $n\in\N$ is s.t. $n>\alpha$.\footnote{Note that the definition 
does not depend on the choice of such an $n$.}
\ed
Finally, we need the following definition.
\bd[Index]
Let $a\in C(\T)$ with $a(t)\neq 0$ for all $t\in\T$, and 
let $c:\T\to\R$ with $c\in C(\T\setminus\{1\})$ be s.t. 
$a=|a|\,\e^{\ii c}$. The index
(or winding number) of $a$ is defined by
\be
\label{index}
\ind(a)
:=\frac{c(1-0)-c(1+0)}{2\pi}.
\ee
\ed
\subsection{Toeplitz operators}
\label{app:Toeplitz}

For $M\in\N$, we denote by  $\ell^2_M(\N)$ 
the space of all square-summable $\C^M$-valued sequences.\footnote{W.r.t. 
the Euclidean norm on $\C^M$.} Moreover, we set
\be
L^\infty_{M\times M}(\T)
:=\{f:\T\to\C^{M\times M}\,|\,  
f_{ij}\in L^\infty(\T)\mbox{ for all $i,j=1,...,M$}\}.
\ee
We then have the following classical result.

\bt[Toeplitz]
\label{thm:Toeplitz}
Let $\{a_x\}_{x\in\Z}\subset\C^{M\times M}$.
The linear operator $A:\dom(A)\subseteq \ell^2_M(\N)\to \ell^2_M(\N)$ 
defined on all $f\in\dom(A)$ with maximal domain $\dom(A)$ by
\be 
Af
:=\left\{\sum_{j=1}^\infty a_{i-j} \,f_j\right\}_{i=1}^\infty,
\ee
is a  
bounded operator on $\ell^2_M(\N)$ if and only if  there exists an 
$a\in L^\infty_{M\times M}(\T)$ s.t., for all $x\in\Z$, it holds
\be
a_x
=\int_{-\pi}^{\pi}\frac{\dd k}{2\pi}\,\,a(\e^{\ii k})\, \e^{-\ii k x}. 
\ee 
\et
\bprf
See B\"ottcher and Silbermann \cite[p.186]{BS99}.
\eprf

We then make the following definition.
\bd[Symbol]
Under the assumptions of Theorem \ref{thm:Toeplitz}, we write 
the Toeplitz operator as $T[a]:=A\in\mL(\ell^2_M(\N))$. It has the matrix 
form
\be
T[a]=\left[  
    \begin{array}{lllll}  
    a_0 & a_{-1} & a_{-2} & ...\\  
    a_1 & a_0 & a_{-1} & ...\\  
    a_2 & a_1 & a_0 &... \\  
    ... & ... & ... & ...\\  
    \end{array}\right].  
\ee
The function $a\in L^\infty_{M\times M}(\T)$ is called the symbol of  
$T[a]$. If $M=1$, the symbol $a\in L^\infty(\T)=L^\infty_{1\times  
1}(\T)$ and the Toeplitz operator $T[a]$ are called scalar, whereas  
for $M>1$ they are called block. 
\ed

Finally, the Toeplitz operators are truncated as follows.
\bd[Finite section]
Let $n\in\N$. The projection $P_n\in\mL(\ell^2_M(\N))$ is defined,
on all $f:=\{x_1,...,x_n,x_{n+1},...\}\in \ell^2_M(\N)$, by
\begin{eqnarray}
\label{def:Pn}
 P_nf=\{x_1,...,x_n,0,0,...\}.
\end{eqnarray}  
Moreover, the truncated Toeplitz matrices  $T_n[a]\in \C^{Mn\times Mn}$ are defined by
\begin{eqnarray*}
T_n[a]
:= P_n T[a] P_n\!\!\upharpoonright_{\ran(P_n)}.
\end{eqnarray*}  
\ed
\subsection{Fisher-Hartwig symbols}
\label{app:FH}

The following theorem summarizes the main results
on the asymptotic behavior of Toeplitz determinants with
Fisher-Hartwig symbols.
\bt[Fisher-Hartwig]
\label{thm:FH}
Let $a\in L^\infty(\T)$ be a restricted Fisher-Hartwig symbol, i.e., for 
all $t\in\T$, the symbol has the form
\be
\label{FH-symbol}
a(t)
=b(t)\prod_{j=1}^m|t-t_j|^{2\alpha_j}\varphi_{\beta_j,t_j}(t),
\ee
and it satisfies the following assumptions:
\begin{enumerate}
\item[(a)] $t_1,...,t_m\in\T$  are pairwise distinct points
\item[(b)] $\alpha_j,\beta_j\in\C$ with $|{\rm Re}(\alpha_j)|<\tfrac12$ and  $|{\rm Re}(\beta_j)|<\tfrac12$ for all $j=1,...,m$
\item[(c)] $b\in L^\infty(\T)\cap B^1_1(\T)$
\item[(d)] $b(t)\neq 0$ for all $t\in\T$
\item[(e)] $\ind(b)=0$
\end{enumerate}
Then, for $n\to\infty$, the Toeplitz determinant has the asymptotic approximation
\begin{eqnarray}
\label{FH}
\det(T_n[a])\sim G(b)^{n} n^Q F(a),
\end{eqnarray}
where the exponential factor and the power law factor are determined 
by
\be
\label{Gb}
G(b)
&:=&\exp\!\left[(\log b)_0\right],\\
\label{Q}
Q
&:=&\sum_{j=1}^m \big( \alpha_j^2-\beta_j^2 \big).
\ee
Here, $f_x$ for $x\in\Z$ denotes the $x$-th Fourier coefficient of the function $f\in L^1(\T)$. Moreover, the constant $F(a)$ is given by
\be
\label{Fa}
F(a)
:=E(b)
\prod_{j=1}^m 
b_+(t_j)^{-(\alpha_j-\beta_j)}
b_-(t_j)^{-(\alpha_j+\beta_j)}
\prod_{j=1}^m
{\rm G}_{\alpha_j,\beta_j}
\!\!\prod_{1\le i\neq j\le m}\!\!\!\bigg(1-\frac{t_i}{t_j}\bigg)^{-(\alpha_i-\beta_i)(\alpha_j+\beta_j)},
\ee
where we define\footnote{Under the assumptions 
on $b$, there exists a logarithm 
$\log b\in W(\T)\cap B_2^{1/2}(\T)$, where $W(\T)$ denotes 
the Wiener algebra, see B\"ottcher and Silbermann \cite[p.123]{BS99}.
Therefore,  $G(b)$, $E(b)$, and $b_\pm(t_j)$ for $j=1,...,m$ are 
well-defined, and the factors on the r.h.s. of \eqref{FH} are independent of the choice of $\log b$.}
\be
\label{Eb}
E(b)
&:=&\exp\!\left(\sum_{l=1}^\infty l(\log b)_l(\log b)_{-l}\right),\\
\label{bpm}
b_\pm(t_j)
&:=&\exp\!\left(\sum_{l=1}^\infty (\log b)_{\pm l} t_j^{\pm l}\right),\\
\label{G1}
{\rm G}_{\alpha_j,\beta_j}
&:=&\frac{{\rm G}(1+\alpha_j+\beta_j){\rm G}(1+\alpha_j-\beta_j)}
{{\rm G}(1+2\alpha_j)}.
\end{eqnarray}
Finally, the function ${\rm G}:\C\to\C$ is the entire Barnes ${\rm G}$-function defined by 
\be
\label{G}
{\rm G}(z+1)
:=(2\pi)^{z/2}\e^{-z(z+1)/2-\gamma_E z^2/2}
\prod_{n=1}^\infty \left[\left(1+\frac{z}{n}\right)\e^{-z+z^2/(2n)}\right],
\ee
where $\gamma_E$ is Euler's constant.
\et
\bprf
See B\"ottcher and Silbermann \cite[p.582]{BS06}.
\eprf
\section{Regularized symbol}  
\label{app:symbol} 

Since we have to care about the behavior in the neighborhood of the 
discontinuities, we write the derivatives of the regularized symbol 
explicitly.
\bl[Regularity]
\label{lem:regsym2}
Let $\T_\pm:=\{t\in\T\,|\, {\rm Im}(t)\neq 0, \, 
\sign({\rm Im}(t))=\pm 1\}$, and set 
$b_{\rP,\T_\pm}:=b_\rP\!\!\!\upharpoonright_{\T_\pm}$.
Then, 
the regularized symbol $b_\rP\in C(\T)$ from Definition 
\ref{def:regsym} has the following properties.
\begin{enumerate}
\item[(a)] $b_{\rP,\T_\pm}\,\in C^\infty(\T_\pm)$ 
\item[(b)] $b_\rP\in C^1(\T)$
\item[(c)] The left and right derivatives $D_\pm b_\rP'(t)$ exist for
 all $t\in\T$, but, for $j=1,2$, we have
\be
\label{DD}
D_-b'_\rP(t_j)
\neq D_+b'_\rP(t_j).
\ee
Moreover, the second derivative is essentially bounded,
\be
\label{norm}
{\|b''_\rP\|}_{L^\infty(\T)}
<\infty.
\ee
\end{enumerate}
\el 
\bprf
Assertion (a) follows from the very form of \eqref{b}. As for 
assertion (b), we find that the one-sided derivatives of $b_\rP$ at 
the points $t_1$ and $t_2$ coincide, and, for $j=1,2$, are given by 
the expression
\be
\label{1st-t12}
D_\pm b_\rP(t_j)
=\frac{1}{2\pi} 
\sqrt{\tau_R(t_j)\tau_L(t_j)}\,
\log\left(\frac{\tau_L(t_1)\tau_R(t_2)}{\tau_R(t_1)\tau_L(t_2)}\right).
\ee
Here, for any $f:\T\to\C$, I used the notation
\be
D_\pm f(t_1)
&:=& \lim_{\eps\to 0^+} \frac{1}{\pm\eps}
(f(\e^{\pm\ii\eps})-f(t_1)),\\
D_\pm f(t_2)
&:=& \lim_{\eps\to 0^+} \frac{1}{\pm\eps}
(f(\e^{\ii(\mp\pi\pm\eps)})-f(t_2)).
\ee
Combining \eqref{1st-t12} with the derivative of $b_{\rP,\T_\pm}$,
we get 
\be
\label{b'}
b'_\rP(\e^{\ii k})
&=& \left(\frac{\tau_L(t_1)\tau_R(t_2)}{\tau_R(t_1)\tau_L(t_2)}\right)^\frac{k}{2\pi}\cdot\nonumber\\
&&\hspace{-1.5cm}\cdot
\begin{cases}
\sqrt{\frac{\tau_R(t_1)}{\tau_L(t_1)}}\,
\tau_L(\e^{\ii k}) 
\left(\frac{1}{2\pi}\log\left(\frac{\tau_L(t_1)\tau_R(t_2)}{\tau_R(t_1)\tau_L(t_2)}\right)
+\beta_L \tilde\tau_L(\e^{\ii k})\sin k\right), 
& \mbox{if\, $0\le k\le \pi$},\\
\sqrt{\frac{\tau_L(t_1)}{\tau_R(t_1)}}\,
\tau_R(\e^{\ii k}) 
\left(\frac{1}{2\pi}\log\left(\frac{\tau_L(t_1)\tau_R(t_2)}{\tau_R(t_1)\tau_L(t_2)}\right)
+\beta_R \tilde\tau_R(\e^{\ii k})\sin k\right), 
& \mbox{if\, $-\pi<k<0$},
\end{cases}
\ee
where, for $\alpha=L,R$, we set $\tilde\tau_\alpha(\e^{\ii k})
:=\tfrac12(1+\tanh[\tfrac12 \beta_\alpha(\lambda+\cos k)])$. Hence,
it follows from \eqref{b'} that $b'_\rP\in C(\T)$. Finally, computing
$b''_{\rP,\T_\pm}$ and the left and right derivatives of $b'_\rP$
at the points $t_1$ and $t_2$, 
\be
D_+ b'_\rP(t_1)
&=&\sqrt{\tau_R(t_1)\tau_L(t_1)}\,
\left(\left[\frac{1}{2\pi} \log\left(\frac{\tau_L(t_1)\tau_R(t_2)}{\tau_R(t_1)\tau_L(t_2)}\right)\right]^2
+\beta_L \tilde\tau_L(t_1)\right),\\
D_- b'_\rP(t_1)
&=&\sqrt{\tau_R(t_1)\tau_L(t_1)}\,
\left(\left[\frac{1}{2\pi} \log\left(\frac{\tau_L(t_1)\tau_R(t_2)}{\tau_R(t_1)\tau_L(t_2)}\right)\right]^2
+\beta_R \tilde\tau_R(t_1)\right),\\
D_+ b'_\rP(t_2)
&=&\sqrt{\tau_R(t_2)\tau_L(t_2)}\,
\left(\left[\frac{1}{2\pi} \log\left(\frac{\tau_L(t_1)\tau_R(t_2)}{\tau_R(t_1)\tau_L(t_2)}\right)\right]^2
-\beta_R \tilde\tau_R(t_2)\right),\\
D_- b'_\rP(t_2)
&=&\sqrt{\tau_R(t_2)\tau_L(t_2)}\,
\left(\left[\frac{1}{2\pi} \log\left(\frac{\tau_L(t_1)\tau_R(t_2)}{\tau_R(t_1)\tau_L(t_2)}\right)\right]^2
-\beta_L \tilde\tau_L(t_2)\right),
\ee
we find, on one hand, that, for $j=1,2$,
\be
D_- b'_\rP(t_j)-D_+ b'_\rP(t_j)
=\sqrt{\tau_R(t_j)\tau_L(t_j)}\,
\left[\beta_R\tilde\tau_R(t_j)-\beta_L\tilde\tau_L(t_j)\right],
\ee
and, on the other hand, we get 
$b''_{\rP}\!\!\upharpoonright_{\overline{\T}_\pm}\in C(\overline{\T}_\pm)$.
Hence, we arrive at assertion (c).
\eprf
  
\end{document}